\documentclass[11pt,a4paper]{article}
\usepackage{amsfonts}
\usepackage{amsmath,amssymb}
\usepackage{epsfig,graphicx}

\input{tcilatex}
\begin{document}

\begin{center}
{\Huge On the lightlike Lorentz violation}

{\Huge \bigskip }

\bigskip

\bigskip

\textbf{\ J.L.~Chkareuli and Z. Kepuladze}

\bigskip

\textit{Institute of Theoretical Physics, Ilia State University, 0162
Tbilisi, Georgia\ \vspace{0pt}\\[0pt]
}

\textit{and} \textit{Andronikashvili} \textit{Institute of Physics, 0177
Tbilisi, Georgia\ }

\bigskip \bigskip \bigskip \bigskip \bigskip

\textbf{Abstract}

\bigskip
\end{center}

We consider the lightlike spontaneous Lorentz invariance violation (SLIV)
appearing through the zero "length-fixing" constraint put on a gauge vector
field, $A_{\mu }A^{\mu }=0$, and discuss its physical consequences in the
framework of a conventional QED and beyond. Again, as in the timelike and
spacelike SLIV cases, $A_{\mu }A^{\mu }=\pm M_{A}^{2}$ ($M_{A}$ is a scale
of SLIV), while this constraint leads to an emergence of the Nambu-Goldstone
modes collected in physical photon, the SLIV itself is still left
unobservable unless gauge invariance in the theory is broken. At the same
time, a crucial difference with the two former cases is that the
asymmetrical vacuum corresponding to the lightlike Lorentz violation appears
infinitely degenerated with all other vacua including the symmetrical one.
We show that this degeneracy can be lifted out by introducing an extra gauge
vector field being sterile with respect to an ordinary matter, though having
some potential couplings with the basic $A_{\mu }$ field. A slight mixing of
them makes the underlying gauge invariance to be partially broken due to
which physical Lorentz invariance occurs broken as well. This may cause a
variety of the Lorentz violating processes some of which are briefly
discussed.

\thispagestyle{empty}\newpage

\section{Introduction}

In some analogy with the nonlinear $\sigma $-model \cite{GL} for pions, one
may think a similar "length-fixing" constraint put on a gauge vector field

\begin{equation}
A_{\mu }A^{\mu }=n^{2}M_{A}^{2}\text{ , \ \ }n^{2}\equiv n_{\nu }n^{\nu
}=\pm 1  \label{const}
\end{equation}%
could lead to a spontaneous Lorentz invariance violation (SLIV) with a
proposed scale $M_{A}$ and a direction determined by an unit vector $n_{\nu
} $. This approach has had a long history, dating back to the model of Nambu 
\cite{nambu} for QED with a nonlinearly realized Lorentz symmetry for the
underlying gauge vector field. Note that a correspondence with the nonlinear 
$\sigma $ model for pions may appear rather suggestive in view of the fact
that pions are the only presently known Nambu-Goldstone (NG) particles whose
theory, chiral dynamics\cite{GL}, is given by the nonlinearly realized
chiral $SU(2)\times SU(2)$ symmetry rather than an ordinary linear $\sigma $
model. The constraint (\ref{const}) means in essence that the vector field $%
A_{\mu }$ develops some constant background value

\begin{equation}
<A_{\mu }(x)>\text{ }=n_{\mu }M_{A}  \label{vev1}
\end{equation}%
and the Lorentz symmetry $SO(1,3)$ breaks down to $SO(3)$ or $SO(1,2)$
depending on the timelike ($n^{2}=+1$) or spacelike ($n^{2}=-1$) nature of
SLIV. This $\sigma $ model type ansatz could in principle provide some
dynamical approach to quantum electrodynamics with photon appearing as
massless NG boson \cite{NJL}. Generally, such an intriguing possibility
first discussed in \cite{bjorken} appears to be very attractive over the
last fifty years having been considered in many different contexts (for some
later developments see \cite{eg,cfn,kraus,jen,kos}).

Returning to the above $\sigma $ model type ansatz, one can see, however,
that in sharp contrast to the nonlinear $\sigma $ model for pions, the
nonlinear QED theory ensures that all Lorentz violating effects turn out to
be unobservable. Indeed, it was shown \cite{nambu, az} that the nonlinear
constraint (\ref{const}) implemented as a supplementary condition into the
standard QED Lagrangian with a charged\ fermion field $\psi (x)$

\begin{equation}
L_{QED}=-\frac{1}{4}F_{\mu \nu }F^{\mu \nu }+\overline{\psi }(i\gamma
\partial +m)\psi -eA_{\mu }\overline{\psi }\gamma ^{\mu }\psi \text{ \ \ }
\label{lag11}
\end{equation}%
(whether it is done by direct substitution or through the Lagrange
multiplier term) leaves physical Lorentz invariance untouched. In fact, this
nonlinear QED contains a plethora of Lorentz and $CPT$ violating couplings
when it is expressed in terms of the emergent NG modes ($a_{\mu }$) stemming
from the corresponding parametrization

\begin{equation}
A_{\mu }=a_{\mu }+n_{\mu }(M_{A}^{2}-n^{2}a^{2})^{\frac{1}{2}}\text{ , \ }%
n_{\mu }a_{\mu }=0\text{ \ \ \ \ (}a^{2}\equiv a_{\mu }a^{\mu }\text{).}
\label{gol}
\end{equation}%
being put and expanded in the Lagrangian (\ref{lag11}) in powers of $%
a^{2}/M_{A}^{2}$. However, the contributions of all these Lorentz violating
couplings to physical processes completely cancel out among themselves, as
has been directly confirmed not only at tree level \cite{nambu}, but also
and in one-loop approximation \cite{az}. Naturally, one may believe that
such a strict cancellation happens due to the underlying gauge invariance in
the QED Lagrangian (\ref{lag11}) in which the supplemental nonlinear
constraint (\ref{const}) appears in fact as a possible gauge choice for the
vector field $A_{\mu }$, while the $S$-matrix remains unaltered under such a
gauge convention\footnote{%
Note that all of this ultimately comes down to whether there exists an
appropriate gauge function related to the constraint (\ref{const}).
Following Nambu's heuristic argument \cite{nambu}, one would think that a
gauge function could exist at least for the timelike constraint case ($%
n^{2}=+1$) when the equation for this function is equivalent to the
well-known Hamilton-Jacobi equation for a massive charged particle moving in
electromagnetic field. However, there is no even such an indirect argument
in a general case, and one has to carry out an actual checking calculation
for every possible Lorentz breaking process.\ }.

So, while the constraint (\ref{const}) leads to an emergence of the zero NG
modes collected in physical photon, the SLIV itself, is shown to be
superficial as it affects only the gauge of the vector potential $A_{\mu }$
at least in the the approximation mentioned above. Later this result was
also confirmed for the spontaneously broken massive QED \cite{kep},
non-Abelian theories \cite{jej}, tensor field gravity \cite{cjt} and other
cases\footnote{%
Many interesting aspects of the constrained QED and other theories has also
been studied in \cite{kbl, e}.}. Also, it has been argued \cite{pb} that the
only way to observe physical Lorentz violation in all the constrained vector
or tensor field theories may appear when the underlying gauge symmetry
happens to be really broken rather than merely constrained by some gauge
condition. Nonetheless, in all cases Lorentz invariance appears in fact
spontaneously broken regardless of whether it is observable or not. Indeed,
this violation is normally hidden in gauge degrees of freedom of a vector
field in a gauge invariant theory. However, when these superfluous degrees
are eliminated by gauge symmetry breaking, the Lorentz violation becomes
observable. In this connection, we will use later on the terms Lorentz
violation and physical Lorentz violation in order to distinguish these two
cases.

Here, we consider the lightlike Lorentz violation appearing through the zero
"length-fixing" constraint put on a gauge vector field 
\begin{equation}
A_{\mu }A^{\mu }=0  \label{constrr}
\end{equation}%
and discuss its physical consequences in the framework of a conventional QED
and beyond. This case has not previously been considered in detail in the
literature and, therefore, we try, in some sense, to fill this gap. We show
that again, as in the timelike and spacelike Lorentz violation (\ref{const}%
), the constraint (\ref{constrr}) put into the QED Lagrangian leads to the
essentially nonlinear theory containing a variety of Lorentz and $CPT$
violating couplings in terms of the corresponding Nambu-Goldstone modes.
Nonetheless, all the Lorentz violating effects, due to the starting QED
gauge invariance, turn out to be strictly cancelled in all the lowest order
processes in the ordinary QED framework. At the same time, a crucial
difference with the former two cases is that the asymmetrical vacuum
corresponding to the lightlike Lorentz violation appears infinitely
degenerated with all other vacua including the symmetrical one. We show that
this degeneracy can be lifted out by introducing an extra gauge vector field
being sterile with respect to an ordinary matter, though having some
potential couplings with the basic $A_{\mu }$ field. A slight mixing of them
makes the underlying gauge invariance to be partially broken due to which
physical Lorentz invariance occurs broken as well. This may cause a variety
of the Lorentz violating processes some of which are briefly discussed.

The paper is organized in the following way. In section 2 we present a
general discussion of the lightlike Lorentz violation. Then in section 3 we
discuss the problem of the vacuum degeneracy being generic for this type of
violation and consider the QED model extension with an extra gauge vector
field that allows to lift out this degeneration. In section 4 we present a
conventional QED accompanied by the lightlike vector field constraint with
all the basic interactions appearing in the emergent framework. We discuss
various low-order physical processes and show that all Lorentz violating
contributions in them are completely cancelled out. In next section 5 we
consider the QED model with two constrained gauge vector fields and describe
their potential mixing mechanism which may lead to physical Lorentz
violation that is then demonstrated by an example of the modified Moller
scattering. And finally, in section 6, we present our conclusion.

\section{General discussion}

Let us take a closer look at the lightlike Lorentz violation provided by the
constraint (\ref{constrr}). In this case, it is clear that the vacuum
expectation value (VEV) of the vector field has to be simultaneously
developed on at least two components, one of which is bound to be the time
one. Accordingly, we can use the following\ parametrization for the vector
field $A_{\mu }$ with the emergent NG modes%
\begin{equation}
A_{\mu }=a_{\mu }+\frac{n_{\mu }}{\sqrt{2}}\sqrt{M_{A}^{2}-n^{2}a^{2}}+\frac{%
n_{\mu }^{\prime }}{\sqrt{2}}\sqrt{M_{A}^{2}-n^{\prime 2}a^{2}}  \label{111}
\end{equation}%
where the Lorentz violation scale is given by the VEV parameter $M_{A}$,
while its direction is determined by the unit vectors $n_{\mu }$ and $n_{\mu
}^{\prime }$\ taken in the form%
\begin{equation}
n_{\mu }=(1,0,0,0)\text{, }n^{2}=1;\text{ }n_{\mu }^{\prime }=(0,0,0,1)\text{%
, }n^{\prime 2}=-1  \label{un}
\end{equation}

For all practical purposes, one can make an obvious simplification for what
follows. Assuming, as \ usual, the Lorentz violation scale is rather high,
we can expand the $A_{\mu }$ field parametrization (\ref{111}) in the series
of the $a_{\mu }^{2}/M_{A}^{2}$ so to have in the lowest approximation%
\begin{equation}
A_{\mu }=a_{\mu }+\frac{n_{\mu }+n_{\mu }^{\prime }}{\sqrt{2}}M_{A}-\frac{%
n_{\mu }-n_{\mu }^{\prime }}{\sqrt{2}}\frac{a^{2}}{2M_{A}}\text{ \ }
\label{Aexp}
\end{equation}%
The vacuum direction is now presented by the pure "lightlike" vector%
\begin{equation}
N_{\mu }=(n_{\mu }+n_{\mu }^{\prime })/\sqrt{2}\text{, }N_{\mu }N^{\mu }=0
\label{N}
\end{equation}%
being orthogonal to the emergent NG modes, as one can immediately confirm
requiring the lightlike constraint (\ref{constrr}) to be fulfilled.

According to directions determined by the unit vectors $n_{\mu }$ and $%
n_{\mu }^{\prime }$ these NG modes are 
\begin{equation}
a_{\mu }=\{a_{1}\text{, \ }a_{2}\text{, }N_{\overline{\mu }}a_{0}\}\text{ \
\ }(\overline{\mu }=0,3)  \label{23}
\end{equation}%
satisfying the orthogonality condition 
\begin{equation}
N_{\mu }a^{\mu }=0\text{ }  \label{o}
\end{equation}%
within the taken accuracy $O(a^{2}/M_{A}^{2})$. Meanwhile, the effective
Higgs component $h$ is given by the equation%
\begin{equation}
h=N_{\mu }A^{\mu }=-a^{2}/2M_{A}  \label{h}
\end{equation}%
which means that, in contrast to the timelike and spacelike SLIV, the Higgs
component does not contain the constant part, while its field contained part
appears rather small.

One can readily confirm that the NG mode spectrum (\ref{23}) corresponds to
the broken generators 
\begin{equation}
\mathcal{J}^{[03]}\text{, \ }(\mathcal{J}^{[01]}-\mathcal{J}^{[31]})/\sqrt{2}%
\text{, \ }(\mathcal{J}^{[02]}-\mathcal{J}^{[32]})/\sqrt{2}  \label{gen}
\end{equation}%
as follows from the projection of the $SO(1,3)$ Lorentz generators%
\begin{equation}
\left( \mathcal{J}^{[\alpha \beta ]}\right) _{\nu }^{\mu }=i\left( \delta
_{\nu }^{\alpha }g^{\beta \mu }-\delta _{\nu }^{\beta }g^{\alpha \mu }\right)
\end{equation}%
onto the vacuum direction given by the VEV vector $N_{\mu }=(n_{\mu }+n_{\mu
}^{\prime })/\sqrt{2}$. The remained subgroup is given then by the generators%
\begin{equation}
\mathcal{E}^{(1)}=(\mathcal{J}^{[01]}+\mathcal{J}^{[31]})/\sqrt{2}\text{, \ }%
\mathcal{E}^{(2)}=(\mathcal{J}^{[02]}+\mathcal{J}^{[32]})/\sqrt{2}\text{, \ }%
\mathcal{E}^{(3)}=\mathcal{J}^{[12]}  \label{gr}
\end{equation}%
having the commutation relations%
\begin{equation}
\lbrack \mathcal{E}^{(1)},\mathcal{E}^{(2)}]=0\text{, \ }[\mathcal{E}^{(3)},%
\mathcal{E}^{(a)}]=-i\epsilon ^{\lbrack 3ab]}\mathcal{E}^{(b)}\text{ \ \ }%
(a=1,2)  \label{com1}
\end{equation}%
as can be directly shown. Remarkably, this algebra formally coincides with
an algebra of the so-called Euclidean $E(2)$ symmetry group which is usually
related to Poincare group (with $\mathcal{E}^{(1)}$ and $\mathcal{E}^{(2)}$
appearing as the translation generators) rather than Lorentz one. However,
just this symmetry subgroup now appears solely due to the lightlike SLIV%
\begin{equation}
SO(1,3)\rightarrow E(2)  \label{e2}
\end{equation}%
as one can readily see from the commutation relations (\ref{com1}).

\section{Asymmetrical vacuum choice}

Let us note that the scale of the lightlike Lorentz violation $M_{A}$
introduced above in the parametrization (\ref{111}) is in fact completely
arbitrary. Therefore, the asymmetrical vacuum is infinitely degenerated in
itself, apart from that it is degenerated with the symmetrical vacuum, $%
\left\langle A_{\mu }\right\rangle =0$, as well. We show now that this
degeneracy can be completely lifted out provided that there is some
interaction coupling in the Lagrangian which includes the vector field $%
A_{\mu }$ itself rather than its square. Such an arrangement would be
possible if there was some extra vector field in the theory with which the $%
A_{\mu }$ field could mix. Then, if this extra field, hereinafter referred
to as $B_{\mu }$, develops the VEV by itself, this may determine the VEV of
the basic $A_{\mu }$ field as well, so that the above degeneracy will be
appear lifted. While generally one might speculate about possible nature of
the extra vector field $B_{\mu }$,\ we propose for uniformity that this is a
massless gauge field as well, which is related to some new local $%
U(1)^{\prime }$ symmetry in the same way as the basic $A_{\mu }$ field is
related to the gauge $U(1)$ symmetry in QED. Moreover, this $B_{\mu }$ field
is also proposed to be constrained by the nonlinear vector field condition%
\begin{equation}
B_{\mu }B^{\mu }=n^{2}M_{B}^{2}\text{ , \ \ }n^{2}=\pm 1  \label{cb}
\end{equation}%
developing, however, timelike or spacelike Lorentz violating VEVs rather
than the lightlike one, as the basic $A_{\mu }$ field does. Whether the $%
B_{\mu }$ field is connected to some (hidden) matter or it is sourceless by
itself, but one way or another, it has to be largely sterile with respect to
an ordinary matter in order not to significantly influence the conventional
QED results. Nonetheless, a slight mixing of the $A_{\mu }$ and $B_{\mu }$
fields in their common potential will make, as we show later, the QED gauge
invariance to be partially broken that leads to the lifting of vacuum
degeneracy, from the one hand, and physical Lorentz violation, from the
other.

In line with the above, we will assume that the vector $A_{\mu }$ and $%
B_{\mu }$\ fields develop their VEVs according to the virtually general
Lagrange-multiplier potential 
\begin{equation}
U(A,B)=\frac{\rho _{A}}{4}(A_{\mu }A^{\mu })^{2}+\frac{\rho _{B}}{4}(B_{\mu
}B^{\mu }-n^{2}M_{B}^{2})^{2}+\frac{\lambda }{2}(A_{\mu }B^{\mu
}-M_{AB}^{2})^{2}  \label{pot}
\end{equation}%
where $\rho _{A}(x)$ and $\rho _{B}(x)$ are introduced as the corresponding
Lagrange multipliers providing the existence of the constraints (\ref%
{constrr}) and (\ref{cb}), while $\lambda $ is an intersecting coupling
constant ($\lambda >0$); $M_{B}$ and $M_{AB}$ are the corresponding
violation scales. This type of the potential with the quadratic Lagrange
multipliers might be considered as the $\sigma $-model type limit of a usual
polynomial potential containing all the possible coupling between the $%
A_{\mu }$ and $B_{\mu }$ fields\footnote{%
This potential can be written in the form 
\begin{eqnarray*}
U(A,B) &=&\frac{\lambda _{A}}{4}\left( A^{2}\right) ^{2}+\frac{\lambda _{B}}{%
4}\left( B^{2}-n^{2}M_{B}^{2}\right) ^{2} \\
&&+(\lambda _{A}^{\prime }A^{2}+\lambda _{B}^{\prime }B^{2})AB+\frac{\lambda
_{AB}}{2}\left( AB-M_{AB}^{2}\right) ^{2}
\end{eqnarray*}%
where the second line terms are proposed to lift the\ light-like vacuum
degeneracy mentioned above ($\lambda _{A,B}$, $\lambda _{A,B}^{\prime }$, $%
\lambda _{AB}$ and $M_{B}^{2}$, $M_{AB}^{2}$ are the corresponding
(positive) coupling constants and mass parameters). One can see that in a
special $\sigma $-model type limit \cite{GL} for the individual vector field
couplings in the potential, $\lambda _{A}\rightarrow \infty $ and $\lambda
_{B}\rightarrow \infty $, the nonlinear constraints $A^{2}=0$ and\ $%
B^{2}=n^{2}M_{B}^{2}$ will appear. So, without loss of generality, the above
potential $U(A,B)$ could eventually include only these constraints being
taken through the Lagrange multiplier terms plus the second line
intersecting terms as true potential terms (the first term in this line can
be included into the second one, once the constraints is implemented into
the Lagrangian). In fact, the final form of this potential is presented in (%
\ref{pot}). Note that such a type of the quadratic Lagrange multiplier
potential has been also discussed in the literature \cite{kbl}).}. One way
or another, one can see that variation of the $U$ under $\rho _{A}$ and $%
\rho _{B}$ immediately leads to the vector field constraints causing
condensation of the vector fields $A_{\mu }$ and $B_{\mu }$. This will be
expressed, as before, through the vector $A_{\mu }$ parametrization (\ref%
{111}) with the lightlike Lorentz violation, while for the vector $B_{\mu }$
we will take the parametrization corresponding to the timelike SLIV 
\begin{equation}
B_{\mu }=b_{\mu }+n_{\mu }\sqrt{M_{B}^{2}-b^{2}}\simeq b_{\mu }+n_{\mu
}M_{B}-n_{\mu }\frac{b^{2}}{2M_{B}}\text{ \ }\ \left( nb=b_{0}=0\right)
\label{Bb}
\end{equation}%
where $n_{\mu }$ is the unit timelike Lorentz vector introduced above in (%
\ref{un}) and $b_{\mu }$ fields correspond, as usual, the emergent NG modes (%
$b^{2}\equiv b_{\mu }b^{\mu }$).

Let us note the $B_{\mu }$ is in fact independent vector field gauging its
own $U(1)^{\prime }$ symmetry and does not interact with the charged\
fermion field $\psi (x)$ in our QED model as the $A_{\mu }$ does. Without
the last term in (\ref{pot}) each vector field carries its own Lorentz type
symmetry, so that only this mixing term reduces this extended symmetry of
the potential to the conventional Lorentz one%
\begin{equation}
SO(1,3)_{A}\times SO(1,3)_{B}\rightarrow SO(1,3)  \label{so}
\end{equation}%
We can see that just the last term in (\ref{pot}) is proposed to lift the\
lightlike vacuum degeneracy mentioned. Indeed, just the intersecting mass
parameter $M_{AB}$ appears to determine the scale of the lightlike violation 
$M_{A}$ in the parametrization (\ref{111}). In fact, the stable minimum of
the potential $U$ (\ref{pot}) is provided by the combined VEV of the vector
fields $A_{\mu }$ and $B_{\mu }$ 
\begin{equation}
\left\langle A_{\mu }B^{\mu }\right\rangle =M_{AB}^{2}  \label{ab}
\end{equation}%
Using then the above parametrizations for the vector fields (\ref{Aexp}) and
(\ref{Bb}) containing their VEV%
\begin{equation}
\left\langle A_{\mu }\right\rangle =(n+n^{\prime })_{\mu }M_{A}/\sqrt{2}%
\text{ , \ }\left\langle B_{\mu }\right\rangle =n_{\mu }M_{B}
\end{equation}%
one readily finds 
\begin{subequations}
\begin{equation}
\left\langle A_{\mu }B^{\mu }\right\rangle =\left\langle A_{\mu
}\right\rangle \left\langle B^{\mu }\right\rangle =(n^{2})M_{A}M_{B}/\sqrt{2}%
=M_{AB}^{2}\text{, }M_{A}/\sqrt{2}=M_{AB}^{2}/M_{B}
\end{equation}%
that, therefore, fixes the scale $M_{A}$ and, as a result, degeneracy of the
lightlike Lorentz violation scale\ appears completely lifted out.

However, for such a lifting some extra price has to be paid, as is shown
below. Actually, putting the vector fields $A_{\mu }$ and $B_{\mu }$
expressions (\ref{Aexp}, \ref{Bb}) into the potential (\ref{pot}) one has in
terms of the NG fields $a_{\mu }$ and $b_{\mu }$ 
\end{subequations}
\begin{equation}
U(a,b)=\frac{\lambda }{2}\left[ (n^{\prime }b)M_{A}/\sqrt{2}%
+(na)M_{B}+O(a^{2},b^{2},ab)\right] ^{2}  \label{pt1}
\end{equation}%
which means that some hyperbolic mixture of the $a$ and $b$ modes acquires
the large mass

\begin{equation}
b^{\prime }=b_{3}\cosh \theta -a_{0}\sinh \theta \text{ , \ }\tanh \theta =%
\sqrt{2}\frac{M_{B}}{M_{A}}  \label{cm}
\end{equation}%
acquires the large mass%
\begin{equation}
m_{b}=\lambda (M_{A}^{2}/2-M_{B}^{2})^{1/2}  \label{gm}
\end{equation}%
while the conjugated combination 
\begin{equation}
a^{\prime }=a_{0}\cosh \theta -b_{3}\sinh \theta  \label{cm'}
\end{equation}%
is left massless.

Let us note that the vacuum direction is not given more by the vector $%
N_{\mu }$ (\ref{N}), as in the one vector field case. Rather, it is given by
a similar mixture of the vectors $N_{\mu }$ and $n_{\mu }$ (\ref{un}) 
\begin{equation}
N_{\mu }^{\theta }=N_{\mu }\cosh \theta -n_{\mu }\sinh \theta  \label{sp}
\end{equation}%
which naturally goes to the old vector $N_{\mu }$ once the vector field
mixing disappears. The vector $N_{\mu }^{\theta }$ properly acts on the
combined NG spectrum involved, as can be directly shown. Indeed, now one has
in total the five massless NG modes, whose collection can be written (using
the constant vectors $N_{\mu }=(1,0,0,1)$ and $n_{\mu }^{\prime }=(0,0,0,1)$
and properly enlarging the low-case indices) in the form 
\begin{equation}
\mathcal{G}_{\mathfrak{m}}=\{a_{1},\text{ }a_{2},\text{ }(N_{\overline{\mu }%
}\cosh \theta -n_{\overline{\mu }}^{\prime }\sinh \theta )a^{\prime },\text{
\ }b_{1^{\prime }},\text{ \ }b_{2^{\prime }}\}\text{ \ \ \ \ }(\mathfrak{m}%
=1,\text{ }2,\text{ }\overline{\mu },\text{ }1^{\prime },\text{ }2^{\prime };%
\text{ \ }\overline{\mu }=0,3\text{ })  \label{gg}
\end{equation}%
So, one can readily confirm that the "orthogonality" condition for them is
well satisfied 
\begin{equation}
N_{\mathfrak{m}}^{\theta }\mathcal{G}_{\mathfrak{m}}=0  \label{00}
\end{equation}%
The NG modes $\mathcal{G}_{\mathfrak{m}}$ correspond to the broken generators%
\begin{equation}
\mathcal{J}^{[31]},\text{ }\mathcal{J}^{[32]},\text{ }\mathcal{J}^{[03]},%
\text{ }\mathcal{J}^{[01]},\text{ }\mathcal{J}^{[02]}  \label{gens}
\end{equation}%
respectively\footnote{%
Noticeably, in contrast to the equality of the $a$ mode components, $%
a_{3}=a_{0}$, stemming from the orthogonality condition\ (\ref{o}) in the
one-vector case, now one has instead $a_{3}=\mathcal{(}\cos \theta -\sin
\theta )a^{\prime }$, while $a_{0}=a^{\prime }\cos \theta +b^{\prime }\sin
\theta $. This follows from comparison of the new mode components $\mathcal{G%
}_{\mathfrak{0}}$ and $\mathcal{G}_{\mathfrak{3}}$ in (\ref{gg}) and the
fact that, unlike the $a_{0}$, the starting mode $a_{3}$ does not mix ($%
a_{3}=\mathcal{G}_{\mathfrak{3}})$. Thus, the mode $a_{3}$ is left a true
zero mode, whereas the $a_{0}$ mode goes to the superposition with a heavy
state $b^{\prime }$.}. Thus, only the generator $\mathcal{J}^{[12]}$\ is
conserved that means the starting Lorentz $SO(1,3)$ symmetry is finally
broken to the plane rotation symmetry $SO(2)$ rather than the $E(2)$
symmetry as in the one-vector field case.

The mass mixing in the zero mode potential (\ref{pt1}) leads to the
significant modification of the $a$ and $b$ mode interactions that may lead
in general to physical Lorentz violation being considered in section 5. For
an extreme mass scale hierarchy $M_{A}>>M_{B}$ this mixing appears very
small, thus leading to an approximate separation of the $A$ and $B$ field
sectors with their own Lorentz symmetries (\ref{so}). In a decoupling limit,
the vacuum vector (\ref{sp}) approaches to the old form, $N_{\mu }^{\theta
}\rightarrow N_{\mu }$ and our QED model contains only massless $a$-modes (%
\ref{23}) corresponding to the pure lightlike SLIV with the remained $%
E(2)_{a}$ symmetry appearing from violation of the Lorentz group $%
SO(1,3)_{A} $ in (\ref{so}). The massive $b_{3}$ and massless $b_{1,2}$
modes are practically sterile in the model possessing the remnant symmetry $%
SO(3)_{b}$ originated from the breaking of the second Lorentz group $%
SO(1,3)_{B}$ in (\ref{so}). Actually, this remnant $SO(3)_{b}$ is in fact
explicitly broken to the $SO(2)_{b}$\ symmetry due to the mass splitting in
the $b$ mode triplet mentioned above.

\section{QED with lightlike vector field constraint}

We start with the constrained QED model with one vector field. However,
since (as was discussed above) the lightlike SLIV can not be consistently
formulated in the one-vector field framework, we then consider in next
section its extension including two vector fields.

\subsection{Basic interactions}

Putting the $A_{\mu }$ parametrization (\ref{Aexp}) into the conventional
QED Lagrangian (\ref{lag11}) and remaining only lowest order interaction
terms we come to the emergent QED model for the zero vector $a_{\mu }$ modes 
\begin{equation}
L_{EQED}=L_{0}+\frac{1}{\sqrt{2}M_{A}}L_{1}+a_{\mu }J^{\mu }+L_{m}
\label{lag}
\end{equation}%
where 
\begin{eqnarray}
L_{0} &=&-\frac{1}{4}f_{\mu \nu }^{2}+\frac{[a_{\mu }(n+n^{\prime })^{\mu
}]^{2}}{2\alpha }\text{ ,}  \notag \\
L_{1} &=&-\frac{a_{\lambda }^{2}}{2}\left( n_{\nu }-n_{\nu }^{\prime
}\right) \left( \partial ^{\mu }f_{\mu \nu }+J_{\nu }\right) \text{ }
\label{uq}
\end{eqnarray}%
and $f_{\mu \nu }$ stands for the conventional stress-tensor of the $a_{\mu
} $ field identified with physical photon\footnote{%
Note that, though the VEV scale $M_{A}$ of the vector field $A_{\mu }$ may
be arbitrary in the one-vector field model we will, however, consider $M_{A}$
as a precisely fixed scale in a view of lifting this degeneracy in the
two-vector field model extension discussed in the previous section.}. Here,
we also included the vector field source current $J_{\mu }$ determined by
some matter field(s) whose kinetic terms are presented in the Lagrangian
part $L_{m}$. For certainty, one can assume this matter field to be some
charged fermion. As to the vector NG modes $a_{\mu }$, they are presented in
the Lagrangians $L_{0}$ and $L_{1}$ in the above including the corresponding
axial gauge fixing term (gauge appearing the limit $\alpha \rightarrow 0$).
\ Their propagator is in fact determined by the lightlike vacuum vector, $%
(n+n^{\prime })_{\mu }$, thus leading to its conventional form 
\begin{equation}
D_{\mu \nu }(a)=\frac{-i}{k^{2}}\left( g_{\mu \nu }-\frac{\left( n_{\mu
}+n_{\mu }^{\prime }\right) k_{\nu }+\left( n_{\nu }+n_{\nu }^{\prime
}\right) k_{\mu }}{(n+n^{\prime })k}\right)  \label{d0a}
\end{equation}%
Note\ that the conventional$\ k_{\mu }k_{\nu }$ part in the bracket, usually
included in a propagator in axial gauge, is now absent because this part is
proportional to $\left( n_{\mu }+n_{\mu }^{\prime }\right) ^{2}$, which is
zero. One can immediately confirm that the propagator $D_{\mu \nu }$
satisfies the both orthogonality conditions 
\begin{equation}
\left( n_{\mu }+n_{\mu }^{\prime }\right) D_{\mu \nu }=0\text{ , \ }k_{\mu
}D_{\mu \nu }=0  \label{c}
\end{equation}%
though the latter is only satisfied on the photon mass shell ($k^{2}=0$).

\subsection{Physical processes}

Actually, the whole picture in the QED with the lightlike constraint (\ref%
{constrr}) is very similar to that of the QED with the spacelike and
timelike constraint cases (\ref{const}). This concerns the form of
propagator (\ref{d0a}) and also all the Lorentz violating couplings as those
of the lowest $1/M_{A}$ order presented above (\ref{uq}), so the higher
order ones which are not shown here to save space. As a consequence, one can
not wait for physical Lorentz violation for the QED with lightlike
constraint inasmuch as it does not occur for the spacelike and timelike
constraint cases, as is strictly confirmed up to one-loop approximation \cite%
{az}.

\subsubsection{\protect\bigskip Compton scattering}

Nonetheless, it seems to be useful to show how the SLIV cancellation
mechanism work taking for an illustration the conventional Compton
scattering with a possible modification following from the Lorentz violating
couplings (\ref{uq}). For modified Compton scattering we have two diagrams
in the $1/M_{A}$ order. The first one is related to the longitudinal photon
exchange between the conventional matter current $J_{\mu }$\ (the term $%
a_{\mu }J_{\mu }$ in the total Lagrangian $L_{tot}$ (\ref{lag}) in the
above)\ and the Lorentz violating three-photon vertex $\Gamma _{\mu \nu \rho
}$ (determined by the new\ Lagrangian part $L_{1}$), while the second one is
the\ contact diagram (following again from the $L_{1}$). So, matrix element
for the first diagram reads as%
\begin{equation}
\emph{M}_{1}=\xi ^{\mu }(q)\xi ^{\nu }(p)\Gamma _{\mu \nu \rho
}(q,p,k)D^{\rho \lambda }(k)J_{\lambda }
\end{equation}%
where $\xi ^{\mu }(q)$ and $\xi ^{\nu }(p)$ are polarization vectors for the
ingoing and outgoing photons, respectively. So, plugging the expressions for
the propagator $D_{\rho \lambda }$ and the three-photon vertex $\Gamma _{\mu
\nu \rho }$, and using the matter current conservation we readily come \ 
\begin{equation}
\emph{M}_{1}=\frac{\xi _{\mu }(q)\xi ^{\mu }(p)}{\sqrt{2}M_{A}}(n-n^{\prime
})_{\lambda }J^{\lambda }
\end{equation}%
which appears to be exactly opposite to the second (contact) diagram matrix
element 
\begin{equation}
\emph{M}_{2}=-\frac{\xi _{\mu }(q)\xi ^{\mu }(p)}{\sqrt{2}M_{A}}(n-n^{\prime
})_{\lambda }J^{\lambda }
\end{equation}%
Therefore, one has a total cancellation for physical Lorentz violation in
the modified Compton scattering 
\begin{equation*}
\emph{M}_{2}+\emph{M}_{1}=0
\end{equation*}

Below, we briefly discuss some other processes appearing in higher orders
and argue that there is no physical Lorentz violation either.

\subsubsection{Other processes}

Many other tree level Lorentz violating processes related to the emergent $%
a_{\mu }$ modes (interacting with each other and the matter fields in the
theory) appear in higher orders in the basic SLIV parameter $1/M_{A}$. They
follow by iteration of couplings presented in our basic Lagrangian (\ref{lag}%
), or from a further expansion of the effective vector field (\ref{111})
inserted into the starting total Lagrangian (\ref{lag11}). However, again as
in the Compton scattering considered above, their amplitudes are essentially
determined by an interrelation between the longitudinal photon exchange
diagrams and the corresponding contact interaction diagrams, which appear to
cancel each other, thus eliminating physical Lorentz violation in the
theory. The full similarity of the high-order Lorentz breaking couplings
with those in the timelike and spacelike QED cases studied earlier \cite{az}
allows to think that such cancellation will remain in one-loop approximation
as well.

So, physical Lorentz invariance in the gauge invariant QED with the
lightlike vector field constraint seems to be eventually maintained. Its
violation might only be possible when the QED model includes some gauge
invariance violating terms. This may appear, for example, if a model
contains the second vector field, as is discussed in next section.

\section{QED with two constrained vector fields}

\subsection{Towards physical Lorentz violation}

As was discussed earlier in section 3, in the two-vector field model where
the vacuum degeneracy is lifted, the emergent photon is influenced by the
degree of mixing with the second vector field. This is depended on the
hierarchy between the VEVs $M_{A}$ and $M_{B}$ of the starting vector fields 
$A_{\mu }$ and $B_{\mu }$ containing the zero modes $a_{\mu }$ and $b_{\mu }$%
, respectively. As follows, the bigger their mixing \ parameter (\ref{cm}) 
\begin{equation}
\delta =\tanh \theta =M_{B}/\sqrt{2}M_{A}
\end{equation}%
is the stronger Lorentz violating effects appear for the $a_{\mu }$ modes
associated with physical photon. So, for a realistic scenario we have to
propose $\delta <<1$. Remarkably, while this mixing is caused by the gauge
noninvariant intersecting term in the potential (\ref{pot}, \ref{pt1}), the
size of the mixing itself does not depend on the corresponding coupling
constant $\lambda $. This constant seems to be naturally small since its
vanishing increases symmetry of the potential, as one can see from (\ref{so}%
). However, even for a very small constant $\lambda $ this mixing could be
the same, and also the massive $b^{\prime }$ mode (\ref{cm}) might be still
heavy due to the proposed large $A_{\mu }$ field VEV $M_{A}$.

In the small $\delta $ approximation and with an accuracy $%
O(a^{2}/M_{A}^{2},b^{2}/M_{B}^{2})$ taken everywhere, one can readily find
that the one-vector field Lagrangian (\ref{lag}) is only disturbed by terms
of the type%
\begin{equation}
\Delta L(a)=-\frac{\delta ^{2}}{2}m_{b}^{2}(na)^{2}-\delta
m_{b}^{2}(na)(n^{\prime }b)  \label{lmix}
\end{equation}%
appearing as a result of mixing of the starting vector $A$ and $B$ fields in
the two field Lagrangian $L(A,B)$. This Lagrangian, apart from their
potential terms in (\ref{pot}), contains the $B$ field kinetic terms as well
that for the emergent NG modes $b_{\mu }$ looks as 
\begin{equation}
L(b)=-\frac{1}{4}b_{\mu \nu }^{2}+\frac{\left( b_{\mu }n^{\mu }\right) ^{2}}{%
2\beta }-\frac{b^{2}}{2M_{B}}n_{\nu }\partial ^{\mu }b_{\mu \nu }-\frac{1}{2}%
m_{b}^{2}(n^{\prime }b)^{2}  \label{lb}
\end{equation}%
where we also included the corresponding axial gauge fixing term (gauge
appearing in the limit $\beta \rightarrow 0$) and the lowest order
interaction term. So, we have now both zero modes $a_{\mu }$ and $b_{\mu }$
in the combined two-vector field theory. In terms of them, the starting $A-B$
mixing provides, as on can see, the large mass to the $(n^{\prime }b)$
component of the $b_{\mu }$ field, while the small one to the $(na)$
component of the $a_{\mu }$ field. Also, there is still left some mixing
term of these components that is presented in (\ref{lmix}). The point is,
however, that due to the gauge invariant $F_{\mu \nu }^{2}$\ structure of
the $a$ and $b$ mode kinetic terms one can not diagonalize these modes
simultaneously both in their mass and kinetic sectors. For this reason, we
remain them unrotated in the Lagrangians (\ref{lmix}, \ref{lb}) treating
their mixing term as some small perturbation provided that the above $\delta 
$\ parameter or VEV ratio $M_{B}/M_{A}$ is really small. Thus, we will
calculate their physical masses through the oscillations provided by this
mixing term. One can readily see that these oscillations practically will
not change the large mass of the $(n^{\prime }b)$ component in (\ref{lb}),
while may significantly disturb the small mass term of the $(na)$ component
in (\ref{lmix}).

As matter of fact, we basically have the same emergent QED picture as in the
one field case which is only disturbed by the small mixing of the $a$ modes
(associated with photons) with the $b$ modes being essentially separated
from physical matter. Nonetheless, this mixing causes, as we see below,
physical Lorentz violation through the $a$ mode propagation. Meanwhile,\ the 
$b$ modes containing the massive $(n^{\prime }b)$ component in the
Lagrangian (\ref{lb}) have the total propagator of the form%
\begin{equation}
ik^{2}D_{\mu \nu }^{(m)}(b)=g_{\mu \nu }-\frac{k^{2}[(k_{\mu }n_{\nu
}+k_{\nu }n_{\mu })(nk)-n^{2}k_{\mu }k_{\nu }]+m_{b}^{2}P_{\mu \nu }}{%
k^{2}-m_{b}^{2}[1-(n^{\prime }k/nk)^{2}]}\frac{1}{(nk)^{2}}  \label{db}
\end{equation}%
where the kinematical tensor\ $P_{\mu \nu }$ given by 
\begin{eqnarray}
P_{\mu \nu } &=&k_{\mu }k_{\nu }+[n_{\mu }^{\prime }(nk)-n_{\mu }(n^{\prime
}k)][n_{\nu }^{\prime }(nk)-n_{\nu }(n^{\prime }k)]  \notag \\
&&+k_{\mu }[n_{\nu }^{\prime }(n^{\prime }k)-n_{\nu }(nk)]+k_{\nu }[n_{\mu
}^{\prime }(n^{\prime }k)-n_{\mu }(nk)]
\end{eqnarray}%
may cause the physical Lorentz violation depending on the mass $m_{b}$
involved. One can readily see that in the massless limit this propagator
goes to the standard vector boson propagator $D_{\mu \nu }(b)$ taken in the
axial gauge $(nb)=0$. However, even in the above extended form it still
satisfies the conventional condition $n^{\mu }$ $D_{\mu \nu }^{(m)}(b)=0$
and also the conditions $n^{\prime \mu }$ $D_{\mu \nu }^{(m)}(b)=0$ and $%
k^{\mu }$ $D_{\mu \nu }^{(m)}(b)=0$ when being taken on the $k^{2}=0$ shell.
Indeed, the same conditions are also satisfied by the polarization vector $%
\xi _{b}^{\mu }(k)$ of the $b_{\mu }$ field.

The total $a$ mode propagator $D_{\mu \nu }^{tot}(a)$, in turn, contains,
apart its own one-mode propagator part $D_{\mu \nu }(a)$ (\ref{d0a}), two
type of contributions stemming from the mass term and mixing term in $\Delta
L(a)$ in (\ref{lmix}), respectively. The first lead exactly to the same type
of propagator $D_{\mu \nu }^{(m)}(a)$ as is the\ $D_{\mu \nu }^{(m)}(b)$ in (%
\ref{db}), though with a proper replacement of the unit vector $n_{\mu }$ by 
$n_{\mu }+n_{\mu }^{\prime }$ and mass $m_{b}^{2}$ by $\delta ^{2}m_{b}^{2}$%
. The second causes the oscillations of the $(na)$ mode into the $(n^{\prime
}b)$ mode and back. Their sum over an infinite number of such oscillations 
\begin{equation}
D_{\mu \nu }^{(osc)}(a)=[n^{\lambda }D_{\lambda \mu }(a)]\left[ \delta
^{2}m_{b}^{4}D^{(m)}(b)\right] [n^{\rho }D_{\rho \nu }(a)]\sum_{r=0}^{\infty
}[D(a)\delta ^{2}m_{b}^{4}D^{(m)}(b)]^{r}  \label{dt}
\end{equation}%
(where $D(a)$ and $D^{(m)}(b)$ stand for $n^{\alpha }n^{\beta }D_{\alpha
\beta }(a)$ and $n^{\prime \alpha }n^{\prime \beta }$ $D_{\alpha \beta
}^{(m)}(b)$, respectively)\ can then be expressed in terms of a geometrical
progression, while its tensorial structure will be\ determined by the end
vectors $n^{\lambda }D_{\lambda \mu }(a)$ and $n^{\rho }D_{\rho \nu }(a)$.$\ 
$Once all these simple, though lengthy, calculations are carried out using
the propagators (\ref{d0a}) and (\ref{db}), the joint contribution, $D_{\mu
\nu }^{(m)}(a)+$ $D_{\mu \nu }^{(osc)}(a)$, leads eventually to the total $a 
$ field propagator \ 
\begin{equation}
ik^{2}D_{\mu \nu }^{tot}(a)=ik^{2}D_{\mu \nu }(a)-\delta ^{2}\frac{m_{b}^{2}%
}{k^{2}-m_{b}^{2}[1-(n^{\prime }k/nk)^{2}]}\frac{P_{\mu }P_{\nu }}{%
[k(n+n^{\prime })]^{2}}  \label{tp}
\end{equation}%
which contains the one-mode propagator part (\ref{d0a}) we are accustomed to
plus small Lorentz violating addition with the kinematical vector $P_{\mu }$ 
\begin{equation}
P_{\mu }=k_{\mu }+n_{\mu }^{\prime }(nk)-n_{\mu }(n^{\prime }k)
\end{equation}

One can readily confirm the total propagator $D_{\mu \nu }^{tot}(a)$
satisfies the same conditions (\ref{c}) as its one-vector field part $D_{\mu
\nu }(a)$. Remarkably, this propagator contains the second pole related to
the massive mode $(n^{\prime }b)$. In its massless limit one comes to the
standard $a$ and $b$ field propagators $D_{\mu \nu }(a)$ and $D_{\mu \nu
}(b) $ taken in axial gauge, and Lorentz invariance is recovered. On the
contrary, for the heavy $(n^{\prime }b)$ mode, when $k^{2}<<m_{b}^{2}$, as
it may actually appear for known physical processes with the longitudinal $a$
mode exchange, the Lorentz violating effects in the leading order does not
really depend on the mass $m_{b}$, but rather only on the mixing $\delta $
parameter. The main thing, however, is that the total propagator $D_{\mu \nu
}^{tot}(a)$, due to the two-vector mixing part in (\ref{tp}), leads to the
nonzero Lorentz breaking result when it is sandwiched between the conserved
currents. A real driving force standing behind this physical Lorentz
violation is indeed a small mixing of the $a$ and $b$ modes.

\subsection{Phenomenological aspects}

\subsubsection{A Moller scattering primer}

In this connection, as some demonstrative example may be considered the
Moller electron-electron scattering whose conventional amplitude being in
the lowest order determined by the longitudinal photon exchange is well
known. Thus, one can use this example to take into account the effects of
the modified propagator (\ref{tp}) for photon field. Normally, one could
consider this scattering in the center of mass system with electrons have
the equal initial energies $E$ and oppositely directed 3-momenta $%
\overrightarrow{p}$. However, in this case, since Lorentz breaking
correction in the propagator is proportional to the transferred energy, one
will have zero effect (more precisely, it happens when there appears $kn=0$
but $kn^{\prime }\neq 0$ in the propagator). Instead, we could take the case
when the interacting electrons have different energies but their 3-momenta
are directed to each other. So, for initial and final momentum we would have
the following configuration 
\begin{eqnarray}
p_{\mu } &=&(E_{p},\text{ }p\overrightarrow{\emph{l}})\text{, \ \ }q_{\mu
}=(E_{q},\text{ }-q\overrightarrow{\emph{l}})\text{\ \ \ }  \notag \\
P_{\mu } &=&(E_{P},\text{ }P\overrightarrow{\emph{l}_{P}})\text{, \ \ }%
Q_{\mu }=(E_{Q},\text{ }Q\overrightarrow{\emph{l}_{Q}})\text{\ \ }  \label{1}
\end{eqnarray}%
respectively, where the unit vectors $\overrightarrow{\emph{l}}$ , $%
\overrightarrow{\emph{l}_{P}}$ and $\overrightarrow{\emph{l}_{Q}}$ stand for
directions of the initial and final 3-momenta involved, while $p$, $q$, $P$
and $Q$ give their values. For high energies one can neglect the electron
mass in the leading order, thus having for the final energy $E_{P}$ of one
of electrons%
\begin{equation}
E_{P}=\frac{2E_{p}E_{q}}{E_{p}(1-\emph{l}\cdot \emph{l}_{P})+E_{q}(1+\emph{l}%
\cdot \emph{l}_{P})}\text{ }  \label{2}
\end{equation}

With the chosen kinematics (\ref{1}, \ref{2}) the total differential cross
section for an elastic $e-e$ scattering can be presented as 
\begin{equation}
\frac{d\sigma ^{tot}}{d\Omega }=\frac{d\sigma }{d\Omega }+\delta ^{2}\frac{%
d\sigma ^{br}}{d\Omega }
\end{equation}%
which contains the Lorentz violating part $d\sigma ^{br}/d\Omega $ as well.
An exact formula for this term is generally too long to be displayed, but
for the case when initial electron energies are close to each other, $%
E_{p}-E_{q}\ll E_{p}+E_{q}$, it is significantly simplified 
\begin{equation}
\frac{d\sigma ^{\prime }}{d\Omega }=-4\alpha ^{2}\frac{\left[ \left( \emph{l}%
\cdot \emph{l}_{P}\right) ^{2}+5\right] \left[ \left( \emph{l}\cdot
n^{\prime }\right) ^{2}+\left( \emph{l}_{P}\cdot n^{\prime }\right) ^{2}%
\right] +4\left( \emph{l}\cdot \emph{l}_{P}\right) \left( \emph{l}\cdot
n^{\prime }\right) \left( \emph{l}_{P}\cdot n^{\prime }\right) }{\left(
E_{p}+E_{q}\right) ^{2}[\left( \emph{l-l}_{P}\right) \cdot n^{\prime
}]^{2}[\left( \emph{l+l}_{P}\right) \cdot n^{\prime }]^{2}}\frac{\left(
E_{p}-E_{q}\right) ^{2}}{\left( E_{p}+E_{q}\right) ^{2}}  \label{ss}
\end{equation}%
Note that its high energy behavior is given by $1/E^{4}$, while the standard
cross section part, $d\sigma /d\Omega $, behaves as $1/E^{2}$. The further
simplification in (\ref{ss}) is related to a choice of a scattering plane
which is still arbitrary. For example, when the starting 3-momentum $%
\overrightarrow{p}$ is taken perpendicular to the unit Lorentz vector $%
n^{\prime }$, which means $\emph{l}\cdot n^{\prime }=0$, this cross section
becomes even much simpler%
\begin{equation}
\frac{d\sigma ^{br}}{d\Omega }=-4\alpha ^{2}\frac{\left( \emph{l}\cdot \emph{%
l}_{P}\right) ^{2}+5}{\left( E_{p}+E_{q}\right) ^{2}(\emph{l}_{P}\cdot
n^{\prime })^{2}}\frac{\left( E_{p}-E_{q}\right) ^{2}}{\left(
E_{p}+E_{q}\right) ^{2}}
\end{equation}%
Many other interesting kinematical configurations are also possible to be
set at an experiment.

A comparison of the Lorentz violating cross section parts taken for
different kinematics may give a clue for an actual physical Lorentz
violation in Moller scattering. One could also try to derive some limit on
the underlying parameter $\delta $ (or ratio of the VEVs of the starting $A$
and $B$ fields) using an accuracy in the experimental determination of the
above cross sections. Particularly, for uncertainty of the $10^{-10}$ order
in it one could have $\delta \sim 10^{-5}$ that would mean that if the VEV
mass scale $M_{A}$ is taken to be of the order of the grand unified scale, $%
10^{15}$ GeV, then the mass $M_{B}$ has to be about $10^{10}$ GeV, that
would be quite admissible from any point of view.

\subsubsection{Sterile Nambu-Goldstone modes}

Basically, in contrast to $a_{\mu }$ field associated with photon, the
massless $b_{\mu }$ field modes (\ref{gg}) stemming from the second vector
field $B_{\mu }$ are staying sterile in the theory since they have not
directly coupled with the matter fermion current $J_{\mu }$ involved.
Nevertheless, one might think that they could interact with the matter
through the $b$ mode oscillation into the $a$ mode (as was discussed in
section 5.1), thus creating some effective vertex $b_{\mu }J^{\mu }$. The
point is, however, that such a vertex with the real $b_{\mu }$\ field turns
to zero because of the mass shell conditions for its propagator $D_{\mu \nu
}^{(m)}(b)$ (\ref{db}) and polarization vector $\xi _{b}^{\mu }(k)$
mentioned above. At the same time, while the single $b$ modes do not
interact with a matter, their pair production is quite possible through both
potential and kinetic sectors of the theory (\ref{pt1}, \ref{lmix}, \ref{lb}%
). However, such processes, say, the $b$ pair production in
electron-positron collisions appears extremely suppressed at laboratory
energies being much smaller than the Lorentz violation scale $M_{A}$.
Particularly, the ratio of the cross section $\sigma (ee\rightarrow bb)$ to
the conventional one, $\sigma (ee\rightarrow \mu \mu )$, appears to be given
approximately by $E^{2}/M_{A}^{2}$ that, say, for energy $E=1$ TeV and scale 
$M_{A}$ taken again around $10^{15}$ GeV is only $10^{-24}$. So, the zero $b$
modes are pretty sterile and the emergent QED is not practically influenced
by an existence of the second vector field $B_{\mu }$ in the theory. The
only real effect of this field is eventually manifested through the Lorentz
violating corrections which appear in the photon propagator.

This and some other issues, where physical Lorentz violation could be
manifested, will be addressed in more details later on.

\section{Conclusion}

We considered the lightlike Lorentz violation appearing through the zero
"length-fixing" constraint put on a gauge vector field, $A_{\mu }A^{\mu }=0$%
, and discuss its physical consequences in the framework of a conventional
QED and beyond. Again, as in the timelike and spacelike Lorentz violation, $%
A_{\mu }A^{\mu }=\pm M_{A}^{2}$, while putting this constraint into the
Lagrangian leads to an emergence of the zero Nambu-Goldstone modes collected
in physical photon, the SLIV itself, is shown to be superficial as it
affects only the gauge of the vector field $A_{\mu }$. Actually, this field
being expressed in terms of the zero NG modes involved leads to the
essentially nonlinear theory containing a variety of Lorentz and $CPT$
violating couplings. Nonetheless, all the Lorentz violating effects, due to
the underlying gauge invariance, turn out to be strictly cancelled in the
lowest order processes, as was demonstrated for some processes in section 4.

At the same time, the lightlike spontaneous Lorentz violation case is
certainly different from spacelike and timelike SLIV cases. Particularly, in
contrast to the former two cases the asymmetrical vacuum corresponding to
the lightlike Lorentz violation, $SO(1,3)\rightarrow $ $E(2)$, appears to be
infinitely degenerated with all other vacua including the symmetrical one.
We show that this degeneracy can be completely lifted by introducing some
extra gauge vector field in the model. As a result, while the timelike and
spacelike Lorentz violation effects are left hidden within gauge degrees of
freedom of a photon, the lightlike one with extra vector field leads to
physical Lorentz violation. This is related to the fact that such an
extension makes gauge invariance broken in the model due to which physical
Lorentz invariance occurs broken as well that was demonstrated by an example
of the modified Moller scattering in the previous section.

The two-vector field mixing mechanism for physical Lorentz violation
proposed here can be also applied to any other type of SLIV, particularly,
to the timelike or spacelike Lorentz violation cases. Some special issue
being crucial for a present consideration is that the second constrained
vector field $B_{\mu }$ related to its own gauge symmetry $U(1)^{\prime }$
is proposed. Whether the massless $B_{\mu }$ field is connected to some
(hidden) matter or it is sourceless by itself, but one way or another, it
has to be largely sterile with respect to an ordinary matter in order not to
significantly influence the conventional QED results (as it naturally
appears). Meanwhile, the corresponding NG modes $b_{\mu }$\ can interact
with the matter through an oscillation into the $a_{\mu }$ modes. Although
the single $b$ modes do not directly interact with a matter, their pair
production turns out quite possible. Whereas such processes, say, the
massless $b$ pair production in electron-positron collisions appears
extremely suppressed at laboratory energies, they might be important in the
early universe.

Another and more conventional scenario could be to make these $b$ mode
rather heavy instead that would seem to be more appropriate from the
phenomenological point of view. Remarkably, this might be reached if the
extra $U(1)^{\prime }$ were spontaneously broken so that the constrained $%
B_{\mu }$ field could acquire the heavy mass, just as was shown in \cite{kep}%
, though in some other context. A very attractive accommodation for this
two-vector mixing mechanism would be then the Standard Model extension
possessing the starting symmetry $SU(2)\times U(1)\times U(1)^{\prime }$
where the present model could be included.

These and related questions are planned to be considered in detail elsewhere.

\section*{ Acknowledgments}

We would like to thank Colin Froggatt, Rabi Mohapatra and Holger Nielsen for
useful discussions and comments on spontaneous Lorentz violation. Z.K.\emph{%
\ }acknowledges financial support from Shota Rustaveli National Science
Foundation (grant \# YS-2016-81)


\begin{thebibliography}{99}
\bibitem{GL} S. Weinberg, \textit{The Quantum Theory of Fields,} v.2,
Cambridge University Press, 2000.

\bibitem{nambu} Y. Nambu, Progr. Theor. Phys. Suppl. Extra 190 (1968).

\bibitem{NJL} Y.~Nambu and G.~Jona-Lasinio, Phys. Rev. \textbf{122} (1961)
345;

J. Goldstone, Nuovo Cimento \textbf{19} (1961) 154.

\bibitem{bjorken} J.D.~Bjorken, Ann. Phys. (N.Y.) \ \textbf{24 }(1963) 174.

\bibitem{eg} T.~Eguchi, Phys. Rev. D \textbf{14 }(1976)\textbf{\ }2755.

\bibitem{cfn} J.L.~Chkareuli, C.D.~Froggatt and H.B.~Nielsen,
Phys.~Rev.~Lett.~\textbf{87 }(2001)\textbf{\ }091601;

Nucl.~Phys.~B \textbf{\ 609 }(2001) 46.

\bibitem{kraus} Per Kraus and E.T. Tomboulis, Phys. Rev. D \textbf{66 }%
(2002) 045015.

\bibitem{jen} A. Jenkins, Phys. Rev. D \textbf{69 }(2004) 105007.

\bibitem{kos} V.A. Kostelecky, Phys. Rev. D \textbf{69 }(2004) 105009; R.
Bluhm and V.A. Kostelecky, Phys. Rev. D 71 (2005) 065008.

\bibitem{az} A.T. Azatov and J.L. Chkareuli, Phys. Rev. D \textbf{73 }(2006)
065026 .

\bibitem{kep} J.L. Chkareuli and Z.R. Kepuladze, Phys. Lett. B \textbf{644}
(2007) 212 .

\bibitem{jej} J.L. Chkareuli and J.G. Jejelava, Phys. Lett. B \textbf{659 }%
(2008) 754.

\bibitem{cjt} J.L. Chkareuli, J.G. Jejelava and G. Tatishvili, Phys. Lett. B
696 (2011) 126;

J.L.~Chkareuli, C.D.~Froggatt and H.B.~Nielsen, Nucl. Phys. B 848 (2011) 498.

\bibitem{kbl} R. Bluhm, S.-H. Fung, and V.A. Kostelecky, Phys. Rev. D 
\textbf{77 }(2008) 065020;

R. Bluhm, N.L. Cage, R. Potting and A. Vrublevskis, Phys. Rev. D \textbf{77 }%
(2008) 12500.

\bibitem{e} O.J. Franca, R. Montemayor and L.F. Urrutia, Phys. Rev. D 85
(2012) 085008;

C.A. Escobar and L.F. Urrutia, Phys. Rev. D 92 (2015) 2, 025042;

C.A. Escobar and L.F. Urrutia, Int. J. Mod. Phys. A 32 (2017) 14, 1750077.7.

\bibitem{pb} J.L. Chkareuli, Z. Kepuladze and G. Tatishvili, Eur. Phys. J. C
55 (2008) 309;

J.L. Chkareuli and Z. Kepuladze, Eur. Phys. J. C 72 (2012) 1954.
\end{thebibliography}
\end{document}